\documentclass[a4paper,fleqn,usenatbib]{mnras}

\usepackage[T1]{fontenc}
\usepackage{ae,aecompl}

\usepackage{graphicx}	
\usepackage{amsmath}	
\usepackage{amssymb}	

\title[Polarized Radiative Transfer]{Polarized Radiative Transfer in Planetary Atmospheres and
the Polarization of Exoplanets}

\author[Bailey, Kedziora-Chudczer \& Bott]{
Jeremy Bailey,$^{1,2}$\thanks{E-mail: j.bailey@unsw.edu.au}
Lucyna Kedziora-Chudczer,$^{1,2}$
Kimberly Bott$^{3,4}$
\\
$^{1}$School of Physics, UNSW Sydney, NSW 2052, Australia\\
$^{2}$Australian Centre for Astrobiology, UNSW Sydney, NSW 2052, Australia\\
$^{3}$University of Washington Astronomy Department, Box 351580, UW Seattle, WA 98195, USA\\ 
$^{4}$NASA Astrobiology Institute Virtual Planetary Laboratory, Box 351580, UW Seattle, WA
98195, USA\\
}

\date{Accepted XXX. Received YYY; in original form ZZZ}

\pubyear{2017}

\begin{document}
\label{firstpage}
\pagerange{\pageref{firstpage}--\pageref{lastpage}}
\maketitle

\begin{abstract} 
We describe the incorporation of polarized radiative transfer into the atmospheric radiative transfer
modelling code \textsc{vstar} (Versatile Software for Transfer of Atmospheric Radiation). Using a vector
discrete-ordinate radiative transfer code we are able to generate maps of radiance and polarization
across the disc of a planet, and integrate over these to get the full-disc polarization. In this way
we are able to obtain disc-resolved, phase-resolved and spectrally-resolved intensity and polarization
for any of the wide range of atmopsheres that can be modelled with \textsc{vstar}. We have tested the code by
reproducing a standard benchmark problem, as well as by comparing with classic calculations of the
polarization phase curves of Venus. We apply the code to modelling the polarization phase curves of
the hot Jupiter system HD 189733b. We find that the highest polarization amplitudes are produced with
optically thick Rayleigh scattering clouds and these would result in a polarization amplitude of 27
ppm for the planetary signal seen in the combined light of the star and planet. A more realistic cloud
model consistent with the observed transmission spectrum results is an amplitude of $\sim$20 ppm.
Decreasing the optical depth of the cloud, or making the cloud particles more absorbing, both have the
effect of increasing the polarization of the reflected light but reducing the amount of reflected
light and hence the observed polarization amplitude.   

\end{abstract}




\section{Introduction}

Polarization is generally ignored in calculations of radiative transfer in studies of the
atmospheres of stars and planets. However, all scattering processes polarize light, so a full
treatment of radiative transfer should take account of polarization. In particular we expect
the reflected light from a planet to be highly polarized when viewed at appropriate angles. 
This polarization has been a useful
tool in studying the atmospheres of solar system objects such as Venus \citep[e.g.][]{lyot29,hansen74a} 
and Titan \citep{tomasko82,west91} and there
is interest in the potential of using polarization to characterize exoplanet atmospheres. This
could be done by observing the small planetary polarization present in the combined light of the
star and planet for an unresolved hot Jupiter system \citep*{seager00}. The effects are small but within the range
of the new generation of high precision polarimeters \citep{hough06,wiktorowicz08,bailey15} 
that can measure polarization at parts per million levels. Polarization measurements could also
potentially be used with future space instruments to detect the presence of liquid water in the atmosphere
\citep{bailey07} or on the surface \citep{zugger10} of Earth-like exoplanets.

There have been a number of calculations of polarization for exoplanets
\citep*[e.g.][]{seager00,stam06,buenzli09,lucas09,madhusudham12,karalidi12,karalidi13}. Most of
these required specialist codes for handling the polarization case, that are different from those
widely used to interpret other exoplanet observations. In this paper we describe the
incorporation of polarized radiative transfer into the highly versatile and thouroughly tested
\textsc{vstar} \citep[Versatile Software for Transfer of Atmospheric Radiation][]{bailey12}
 radiative transfer code.
With this code we are able to use a single model to predict the emission spectrum and transmission
spectrum as well as the reflected light and polarization phase curves. 

\textsc{vstar} has been applied to the
atmospheres of the Earth \citep{bailey07a,cotton14}, Venus
\citep{bailey08,bailey09,cotton12,chamberlain13}, Jupiter \citep{chudczer11}, Titan
\citep{bailey11}, brown dwarfs \citep{yurchenko14} and
extrasolar planets \citep{zhou13,zhou15,mancini16}. It
uses a line-by-line approach to molecular absorption combined with a full treatment of
scattering by molecules, clouds and aerosols. Its rigorous approach to radiative transfer has
been tested by comparison with standard benchmark problems, and it's results have been verified by comparison with
both Earth atmosphere and stellar atmosphere reference codes \citep{bailey12}. 

In section 2 we describe the requirements for polarized
radiative transfer, and how we have incorporated this into \textsc{vstar}, making use of a comprehensive
vector radiative transfer solver \citep{spurr06}. In section 3 we describe a test of the
resulting code against a standard benchmark problem \citep{garcia89} in polarized radiative transfer. In section
4 we describe how to generate polarization images across a planetary disc and to obtain the disc-integrated 
polarization at any phase angle. As a test of the
code we compare our model for the polarization phase curves of
Venus with the results of \citet{hansen74a} in section 5. In section 6 we apply the same methods to the polarization of the exoplanet HD 189733b,
first constructing a model for the planet's atmosphere that is consistent with transit and eclipse
observations from space, that is then used to calculate the expected polarization
phase curve.

\section{Approach}

Our approach is similar to that used in the standard version of \textsc{vstar} described by \citet{bailey12}.
Normally, in the non-polarization case we use the \textsc{disort} package \citep{stamnes88} which performs the
radiative transfer solution using the discrete-ordinate method. The radiative transfer equation 
solved in this case has the form:

\begin{equation}
\mu \frac{dI_{\nu}(\tau,\mu,\phi)}{d \tau} = I_{\nu} (\tau, \mu, \phi) - S_{\nu}
(\tau, \mu, \phi)
\label{scalar_rt}
\end{equation}

\noindent where $I_{\nu}$ is the monochromatic radiance (sometimes referred to as intensity or specific
intensity) at frequency $\nu$, and is a function
of optical depth $\tau$, and direction $\mu$, $\phi$, where $\mu$ is the cosine of
the zenith angle, and $\phi$ is the azimuthal angle. The source function $S_{\nu}$ is
given by:

\begin{align}
S_{\nu} (\tau,\mu,\phi) & = \frac{\varpi(\tau)}{4 \pi} \int_0^{2 \pi} \!
\int_{-1}^{1} \! P(\mu,\phi; \mu', \phi') I_{\nu} (\tau, \mu', \phi') d \mu' d \phi' 
\nonumber \\
  &  \mbox{} + (1 - \varpi(\tau)) B_{\nu} (T)  \label{scalar_source} \\
  &  \mbox{} + \frac{\varpi(\tau) F_{\nu}}{4 \pi} P(\mu,\phi; \mu_0, \phi_0) \exp{(-\tau / \mu_0)}
  \nonumber
\end{align}

\noindent where the first term describes scattering of radiation into the beam from other
directions according to single scattering albedo $\varpi$ and phase function $P(\mu,\phi; \mu', \phi')$, the second term
allows for thermal emission, with $B_{\nu} (T)$ being the Planck function and the third term corresponds to direct illumination of the atmosphere by
an external source with flux $\mu_0 F_{\nu}$ and direction $\mu_0, \phi_0$ (e.g. the Sun).  

In order to solve this equation we need to provide, for each atmospheric layer at each wavelength,
the temperature $T$, the vertical optical depth $\Delta\tau$, the single scattering albedo $\varpi$ (i.e. what fraction
of the optical depth is due to scattering rather than absorption), and the phase function $P(\Theta)$
that describes the angular distribution of scattered light. These quantities are obtained by
combining the contributions of line and continuum absorbers and scattering from molecules and 
particles (clouds and aerosols) as described in detail in \cite{bailey12}.

\subsection{The Vector Radiative Transfer Equation}

To take account of polarization we need to replace equations \ref{scalar_rt} and \ref{scalar_source}
with the polarized, or vector, radiative transfer equation that has the form: 

\begin{equation}
\mu \frac{d\mathbf{I}_{\nu}(\tau,\mu,\phi)}{d \tau} = \mathbf{I}_{\nu} (\tau, \mu, \phi) - \mathbf{S}_{\nu}
(\tau, \mu, \phi)
\label{vector_rt}
\end{equation}

\noindent where $\mathbf{I}_{\nu}$ is the Stokes vector with components $\{I, Q, U, V\}$ describing polarized
light. The source function is now:

\begin{align}
\mathbf{S}_{\nu} (\tau,\mu,\phi) & = \frac{\varpi(\tau)}{4 \pi} \int_0^{2 \pi} \!
\int_{-1}^{1} \! \mathbf{\Pi}(\mu,\phi; \mu', \phi') \mathbf{I}_{\nu} (\tau, \mu', \phi') d \mu' d \phi' 
\nonumber \\
  &  \mbox{} + (1 - \varpi(\tau)) B_{\nu} (T) \mathbf{U}  \label{vector_source} \\
  &  \mbox{} + \frac{\varpi(\tau) F_{\nu}}{4 \pi} \mathbf{\Pi}(\mu,\phi; \mu_0, \phi_0) \exp{(-\tau / \mu_0)} \mathbf{U}
  \nonumber
\end{align}

\noindent where $\mathbf{U}$ is the unit unpolarized Stokes vector $\{1, 0, 0, 0\}$, and $\mathbf{\Pi}$ is the phase
matrix which replaces the phase function used in the scalar case, and describes both the angular
distribuiton of scattering and the polarizing effects of scattering. This version of the equation 
is derived under the assumption that the medium is ``macroscopically isotropic and
symmetric'' \citep{mishchenko02}, which is the case if the scattering aerosols 
are either spherical particles, or randomly-oriented aspherical particles.
It would not be valid for particles that had a preferred orientation in the atmosphere. We also 
assume that the illuminating source is unpolarized.

\subsection{The Normalised Scattering Matrix}

It can be seen that the additional information we need to provide for the polarization case is the 4 x 4
phase matrix $\mathbf{\Pi}$. In practice we use the normalised scattering matrix $\mathbf{F}$. The two differ in the coordinate
system used, with $\mathbf{\Pi}$ being defined in the coordinate system of the atmosphere, while
$\mathbf{F}$ is defined relative to the scattering plane (the plane containing the source, scatterer and
observer). The phase matrix can be derived from the
scattering matrix by applying a coordinate rotation as described in \citet{mishchenko02}.

Under our assumptions of a macroscopically isotropic and symmetric medium the scattering matrix
depends only on the angle $\Theta$ between the incident and scattered light, and has symmetries such
that it can be expressed as follows:

\begin{equation}
\mathbf{F}(\Theta) = \begin{bmatrix} a_1(\Theta) & b_1(\Theta) & 0 & 0 \\
b_1(\Theta) & a_2(\Theta) & 0 & 0 \\
0 & 0 & a_3(\Theta) & b_2(\Theta) \\
0 & 0 & -b_2(\Theta) & a_4(\Theta) 
\end{bmatrix}
\end{equation}

Thus there are six independent elements $(a_1, a_2, a_3, a_4, b_1, b_2)$ in the scattering matrix, 
and for the special case of spherically symmetric scattering particles, in addition $a_1 = a_2$ and
$a_3 = a_4$, so there are only four independent elements.

The scattering matrix is normalised such that:

\begin{equation}
\frac{1}{2} \int_0^\pi \! a_1(\Theta) \sin{\Theta} d\Theta = 1
\end{equation}

and it can be seen that $a_1(\Theta)$ is equivalent to the phase function $P(\Theta)$ used in the
scalar radiative transfer equation.

\subsection{Expansion in Generalised Spherical Functions}

In the scalar case it is common to express the phase function as an expansion in Legendre polynomials

\begin{equation}
P(\Theta) = \sum_{l=0}^{l_{\mbox{\tiny max}}}{ \beta_l P_l(\cos{\Theta})}
\label{eqn_mom}
\end{equation}

where the coefficients $\beta_l$ are referred to as the moments of the phase function, and we can
choose the value of $l_{\mbox{\tiny max}}$ depending on the accuracy required.

An equivalent approach for the scattering matrix elements required in the polarized case is to expand
in generalised spherical functions \citep{rooij84,mishchenko02} as follows:

\begin{equation}
a_1(\Theta) = \sum_{l=0}^{l_{\mbox{\tiny max}}}{ \beta_l P_{00}^l (\cos{\Theta})}
\label{eqn_greek1}
\end{equation}

\begin{equation}
a_2(\Theta) + a_3(\Theta) = \sum_{l=0}^{l_{\mbox{\tiny max}}}{(\alpha_l+\zeta_l) P_{22}^l (\cos{\Theta})}
\end{equation}

\begin{equation}
a_2(\Theta) - a_3(\Theta) = \sum_{l=0}^{l_{\mbox{\tiny max}}}{(\alpha_l-\zeta_l) P_{2,-2}^l (\cos{\Theta})}
\end{equation}

\begin{equation}
a_4(\Theta) = \sum_{l=0}^{l_{\mbox{\tiny max}}}{ \delta_l P_{00}^l (\cos{\Theta})}
\end{equation}

\begin{equation}
b_1(\Theta) = \sum_{l=0}^{l_{\mbox{\tiny max}}}{ \gamma_l P_{02}^l (\cos{\Theta})}
\end{equation}

\begin{equation}
b_2(\Theta) = - \sum_{l=0}^{l_{\mbox{\tiny max}}}{ \epsilon_l P_{02}^l (\cos{\Theta})}
\label{eqn_greek6}
\end{equation}

The definition and properties of the generalised spherical functions $P_{mn}^l$ can be found in
appendix B of \citet{mishchenko02}. The function $P_{00}^l$ is identical to the Legendre polynomial
$P_l$ so equation \ref{eqn_greek1} is the same as equation \ref{eqn_mom}.

The notation shown in equations \ref{eqn_greek1} to \ref{eqn_greek6} using the first six greek letters
($\alpha_l, \beta_l, \gamma_l, \delta_l, \epsilon_l, \zeta_l$) is a common covention \citep[e.g.
][]{vestrucci84,garcia89} and these are sometimes referred to as the {\it Greek constants} or {\it Greek
coefficients} (although in the case of an atmosphere these are not constants, but functions of
optical depth and wavelength).

An alternate notation \citep[e.g ][]{rooij84,mishchenko02} uses ($\alpha_1^l, \alpha_2^l, \alpha_3^l,
\alpha_4^l, \beta_1^l, \beta_2^l$) for the expansion coefficients with

\begin{align*}
\alpha_1^l &= \beta_l   &   \alpha_2^l &= \alpha_l  &  \alpha_3^l &= \zeta_l  \\
\alpha_4^l &= \delta_l  &    \beta_1^l &= \gamma_l  &  \beta_2^l  &= \epsilon_l   \\
\end{align*}

\subsection{Vector Radiative Transfer Solvers}

To solve the vector radiative transfer equation as described above we use the Vector Linearised
Discrete Ordinate Radiative Transfer (\textsc{vlidort}) code of \citet{spurr06} that uses the
discrete-ordinate method. We used \textsc{vlidort} version 2.5, the most recent
\textsc{fortran} 77 version.  

The discrete-ordinate method replaces the integral over $\mu'$ that appears
in the source function scattering term (equations \ref{scalar_source} or \ref{vector_source}) by a sum
using Gauss's quadrature formula. This leads to an equation that can be solved using matrix methods.
Typically a double Gauss scheme is used in which a separate set of quadrature angles are used for $-1
< \mu < 0$ and $0 < \mu < 1$. By increasing the number of quadrature angles (or streams) the accuracy
of the angular representation of the radiance field can be improved, at the cost of increased
computation time.

The \textsc{vlidort} code is a comprehensive implementation of the discrete-ordinate method for polarized
radiative transfer. It has been widely used for applications in Earth atmosphere observation
\citep[e.g.][]{cuesta13,hammer16}. However, it is equally suitable for applications in astronomy, and
\citet{cotton17a} used it for modelling stellar atmosphere polarization. In that work stellar
atmosphere polarization predictions using \textsc{vlidort} were verified against previous results using different
methods from \citet{harrington15} and \citet{sonneborn82}. 

We also tried, as an alternative, the \textsc{rt3} code of \citet{evans91} that uses the adding-doubling
method. This method, is a development of the doubling method, originally introduced by
Van de Hulst  \citep[e.g.][]{vandehulst68} and extended to account for polarization by
\citet{hansen71}. We found results from this code to be less accurate in dealing with the benchmark problem
described below, and the code was less comprehensive in its capabilities, so we did not continue with it for the later stages of the
project.

\subsection{Rayleigh Scattering from Molecules}

The scattering matrix expansion coefficients for Rayleigh scattering are given in table \ref{tab_rayleigh} in terms of
the depolarization factor $\rho$ \citep{spurr06}. In the absence of depolarization they become ($\beta_0 = 1.0, \beta_2 =
0.5, \alpha_2 = 3.0, \gamma_2 = -\sqrt{6}/2$ and $\delta_1 = 1.5$).

The Rayleigh scattering cross section ($\sigma$) is derived from the refractive index of the gas
using \citep{hansen74}:

\begin{equation}
\sigma = \frac{8 \pi^3 (n(\lambda)^2 - 1)^2}{3 \lambda^4 N^2} \frac{6 + 3\rho}{6 - 7\rho} 
\end{equation}

\noindent where $n(\lambda)$ is the wavelength dependent refractive index, $\lambda$ is wavelength and N is the 
number density (molecules
cm$^{-3}$) of the gas measured at the same temperature and pressure as the 
refractive index.

\begin{table}
	\centering
	\caption{Scattering matrix expansion coefficients for Rayleigh scattering}
	\label{tab_rayleigh}
	\begin{tabular}{lcccccc} 
		\hline
		 & $\alpha_l$ & $\beta_l$ & $\gamma_l$ & $\delta_l$ & $\epsilon_l$ & $\zeta_l$ \\
		\hline
		$l = 0$  & 0 & 1 & 0 & 0 & 0 & 0 \\
		$l = 1$  & 0 & 0 & 0 & $\frac{3 (1 - 2 \rho)}{2 + \rho}$ & 0 & 0 \\
		$l = 2$  & $\frac{6 (1 - \rho)}{2 + \rho}$ & $\frac{1 - \rho}{2 + \rho}$ &
		$-\frac{\sqrt{6}(1 - \rho)}{2 + \rho}$ & 0 & 0 & 0 \\
		\hline
	\end{tabular}
\end{table}

\begin{table*}
\centering
\caption{Comparison with ``L = 13'' benchmark test \citep{garcia89}}
\label{tab_l13}
\begin{tabular}{lllll}
\hline
Stokes & $\mu$ & Garcia/Siewert & \textsc{vstar/vlidort} & Difference \\
\hline
I    &   $-$1.0  &   0.0549566  &  0.0549567 &  +0.0000001   \\
     &   $-$0.9  &   0.0904913  &  0.0904913 &  +0.0000000   \\
     &   $-$0.8  &   0.125601   &  0.125601  &  +0.000000    \\
     &   $-$0.7  &   0.167810   &  0.167810  &  +0.000000    \\
     &   $-$0.6  &   0.219343   &  0.219343  &  +0.000000    \\
     &   $-$0.5  &   0.282944   &  0.282944  &  +0.000000    \\
     &   $-$0.4  &   0.362682   &  0.362682  &  +0.000000    \\
     &   $-$0.3  &   0.465232   &  0.465233  &  +0.000001    \\
     &   $-$0.2  &   0.602877   &  0.602878  &  +0.000001    \\
     &   $-$0.1  &   0.802239   &  0.802239  &  +0.000000    \\
     &   $-$0.0  &   1.11633    &  1.11633   &  +0.00000     \\
     \\
Q    &   $-$1.0 &   $-$0.0216088 &  $-$0.0216889 & $-$0.0000001   \\
     &   $-$0.9 &   $-$0.0325810 &  $-$0.0325810 &    +0.0000000  \\
     &   $-$0.8 &   $-$0.0350476 &  $-$0.0350477 & $-$0.0000001   \\
     &   $-$0.7 &   $-$0.0349497 &  $-$0.0349497 &    +0.0000000  \\
     &   $-$0.6 &   $-$0.0327675 &  $-$0.0327675 &    +0.0000000  \\
     &   $-$0.5 &   $-$0.0286635 &  $-$0.0286636 & $-$0.0000001   \\
     &   $-$0.4 &   $-$0.0227539 &  $-$0.0227539 &    +0.0000000  \\
     &   $-$0.3 &   $-$0.0152402 &  $-$0.0152402 &    +0.0000000  \\
     &   $-$0.2 &   $-$0.00664174 & $-$0.00664175 & $-$0.00000001 \\
     &   $-$0.1 &   $-$0.00143775 & $-$0.00143773 &   +0.00000002 \\
     &   $-$0.0 &   $-$0.00080295 & $-$0.00080289 &   +0.00000006 \\
\hline
\end{tabular}
\end{table*}

\subsection{Scattering from Particles}

The scattering properties of particles (e.g. clouds and aerosols) can be calcuated using Lorenz-Mie theory. \textsc{vstar} uses 
a Lorenz-Mie scattering code from \citet{mishchenko02} that models scattering from a size distribution of spherical
particles. This code returns the extinction and scattering cross sections and hence the single scattering
albedo. The code also calculates
the scattering matrix expansion coefficients needed as input to the radiative transfer solvers. 

The number of coefficients needed in equations \ref{eqn_greek1} to \ref{eqn_greek6} to accurately represent
the scattering matrix generally increases with the size parameter $x = 2 \pi a/\lambda$ where $a$ is the particle
radius, and $\lambda$ is the wavelength. For $x <$ 1 we are close to the Rayleigh regime and only a small number of
expansion coefficients are needed. For large $x$ the number of terms can become much higher. This in turn determines
the number of streams needed in the discrete-ordinate radiative transfer solution to accurately represent the
angular dependence of the radiation field. The number of streams in each hemisphere should be at least half the number of
expansion coefficients for best accuracy.

\subsection{Integration into VSTAR}

\textsc{vstar} implements a model of an atmosphere with a user-defined number of layers each of which has a specified
composition in terms of line and continuum absorbers (atoms, molecules and ions) and potentially a
number of aerosol components. The gas phase composition of the layers can be prescribed, or can be
obtained from a chemical equilibrium model given specified elemental abundances.
\textsc{vstar} combines the optical properties of these various components with appropriate weighting to
determine for each layer and at each wavelength the optical properties required for the radiative
transfer equation solver \citep[see][for a detailed description]{bailey12}. The only change required for
the polarization case is to replace the scalar radiative transfer solver \textsc{disort} with the vector solver
\textsc{vlidort} and calculate the scattering matrix expansion
coefficients described above. These are calculated for each layer of the atmosphere at each
wavelength being modelled. If a layer of the atmosphere contains more than one scattering component (i.e.
different cloud or aerosol types) the optical depths are summed and the expansion coefficents for the components are
averaged, weighting according to each components' contribution to the optical depth.

\section{Benchmark Test}

The benchmark test we consider is the ``L = 13'' problem of \citet{garcia89}. In this problem we model
scattering of light with wavelength 0.951 $\mu$m from a gamma distribution of spherical particles with
effective radius $r_{\mbox{\small eff}}$ = 0.2 $\mu$m, effective variance $v_{\mbox{\small eff}}$ = 0.07 and refractive index
1.44.  The single scattering albedo is $\varpi$ = 0.99. The particles are contained in a uniform slab
with optical depth $\tau$ = 1, and illuminated by a source at $\mu_0$ = 0.2. The slab has a Lambert
surface with albedo 0.1 at its base. \citet{garcia89} tabulate results for this problem accurate to
six significant figures. In table \ref{tab_l13} we present our results for this problem implemented in
\textsc{vstar}. The results presented are for the I and Q Stokes parameters as seen at the top of the
atmosphere for values of $\mu$ from 0.0 to -1.0 for the case $\phi - \phi_0 = 0.0$. The \textsc{vstar}
calculations used 30 streams in each hemisphere.

For \textsc{vstar} using \textsc{vlidort} the results are in most cases in agreement with the benchmark results to
$\pm$1 in the sixth significant figure. Good agreement is also found for other results tabulated by
\citet{garcia89} for this problem using different relative azimuths and optical depths. We note that
while \textsc{vlidort} has previously been reported to reproduce this benchmark result, our test, in addition,
validates the correct calculation of the scattering matrix expansion from the basic particle
properties in \textsc{vstar}.

\section{Polarization across a Planetary Disc}

\subsection{Method}

Up to now the models we have considered describe polarization at a single point on a planetary surface that can be
represented by a locally plane-parallel atmosphere. However, in many cases we are interested in the
distribution of polarization across the visible disc of the planet, or in the integrated polarization
from the whole disc. The latter is particularly the case when considering polarization of exoplanets
where the planet itself is unresolved.

One approach to integrating over the disc is to use a Gaussian quadrature scheme \citep{horak50,hansen74a}.
\citet{stam06} describe an efficient way of integrating over the disc of a
horizontally homogenous planet. These approaches work well if only the disc-integrated results are needed.
However, viewing the distribution of polarization across the disc can be instructive in helping to understand
the polarization features seen in the integrated signal. There are also polarization effects such as limb polarization
\citep{schmid06} that, due to symmetry, are lost in the integrated signal.  We therefore follow a different approach which allows us to 
derive from a model an image of the distribution of polarization across the disc, as well as the integrated
polarization properties.

We start by considering an orthographic projection of the spherical planet onto a plane in which we
define a cartesian coordinate  system $x, y$ with coordinates normalised so that the projected radius
of the planet is one. An orthographic projection is  equivalent to the view of the planet from a large
distance. We introduce the coordinates  $\zeta$ and $\eta$, that are the longitude and colatitude of a point 
on the planet related to $x, y$ via:

\begin{equation}
\eta = \arccos{y}
\label{eqn_ortho}
\end{equation}

\begin{equation}
\zeta = \arctan{(\frac{x}{\sqrt{1-r^2}})}
\end{equation}

\noindent where $r = \sqrt(x^2 + y^2)$. 

In this coordinate system the ``equator'' is the plane containing the
illuminating star and the observer, and the zero of longitude is at a point immediately below the
observer. 

We can now obtain the zenith angles $\mu$, $\mu_0$ and the relative azimuth $\phi - \phi_0$ needed for
a radiative transfer solution  for any phase angle $\alpha$ using \citep{horak50}:

\begin{equation}
\mu_0 = \sin{\eta} \cos{(\zeta - \alpha)}
\end{equation}

\begin{equation}
\mu = \sin{\eta} \cos{\zeta}
\end{equation}

\begin{equation}
\cos{(\phi - \phi_0)} = \frac{\mu \mu_0 - \cos{\alpha}}{\sqrt{1-\mu^2} \sqrt{1-\mu_0^2}}
\end{equation}

\begin{equation}
\sin{(\phi - \phi_0)} = \frac{\sin{\alpha} \cos{\eta}}{\sqrt{1-\mu^2} \sqrt{1-\mu_0^2}}  
\label{eqn_phi}
\end{equation}

We also need the angle $\kappa$ between the local meridian plane at a point on the planet and the planetary
scattering plane \citep{stam06}. Polarization vectors derived from a local radiative transfer solution need rotating
through this angle to give vectors in our $x,y$ coordinate system.

\begin{equation}
\sin{\kappa} = \frac{\cos{\eta}}{\sqrt{1-\mu^2}}
\end{equation}

\begin{equation}
\cos{\kappa} = \frac{\sin{\eta}\sin{\zeta}}{\sqrt{1-\mu^2}}
\end{equation}

Note that the angles $\zeta$, $\phi-\phi_0$ and $\kappa$ can range through all four quadrants so must be calcuated 
in a way that places them in the correct quadrant (e.g. using the atan2 function).

We apply this model by generating a grid of uniformly spaced ``pixel centres'' across the disc of the
planet in the  x,y plane. For each point that is within the illuminated disc we use equations
\ref{eqn_ortho} to \ref{eqn_phi}  to calculate the corresponding values of $\mu$, $\mu_0$, and
$\phi-\phi_0$. We then perform the vector radiative transfer solution for the atmosphere for this set
of geometries with \textsc{vlidort}. Using \textsc{vlidort} it is possible to calculate many values of $\mu$, $\mu_0$
and $\phi-\phi_0$ in a single computation. This typically allows $\sim$ 100 pixels of a disc image to
be computed simultaneously. Time can also be saved because of the symmetry of the situation.
Corresponding pixels in the northern and southern hemisphere are equivalent, apart from a change in
the sign of $U$, and only need to be computed once.

The resulting Stokes vectors can then be converted to Stokes vectors in the x,y coordinate system by
rotating $Q, U$ through $2 \kappa$. The resulting data can be plotted as an image of the intensity and
polarization vectors of the disc of the planet.

To obtain the disc integrated properties it is simply necessary to integrate the Stokes vector
components over the area of the disc as follows using

\begin{equation}
Q_{int} = \int_{-1}^{+1} \! \int_{x1}^{x2} \! Q(x,y) dx dy
\end{equation}

\noindent for the $Q$ Stokes parameter and similar expressions for $I,U$ and $V$.

The limits on the x integration are (for $\alpha > 0$):

\begin{equation}
x_1 = -\sqrt{(1 - y^2)} \sin{(\alpha + \pi/2)}
\end{equation}

\begin{equation}
x_2 = \sqrt{(1 - y^2)}
\end{equation}

To integrate over the grid of equally spaced points we use closed Newton-Cotes formulae of appropriate
order for up to four points and extended Newton-Cotes formulae for larger numbers of points
\citep{press92} to integrate between the first and last grid points on the illuminated disc. We then
add a correction to account for the additional distance from the edges of the disc to the first grid
point at each end using the trapezoidal rule.

\subsection{Test of Integration Accuracy}

\begin{figure}
\includegraphics[width=\columnwidth]{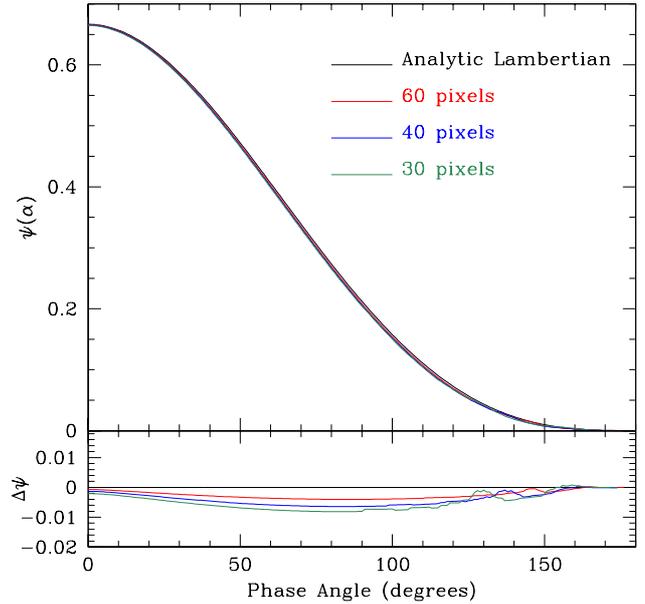}
\caption{Planetary phase function $\psi(\alpha)$ for a Lambertian reflecting spherical planet from the analytic formula, compared
with the results of our integration method. The integration has been carried out over a grid of pixels with either 30, 40
or 60 pixels across the equator of the planet. The bottom panel shows the difference from the analytic formula.}
\label{fig_phasefunc}
\end{figure}

To test the accuracy of our integration approach we have run the code on a model that consists of simply a
Lambertian reflector with surface albedo of one. A spherical planet with a Lambertian surface has an analytic form
\citep{russell16} for the planetary phase
function $\psi(\alpha)$, the variation of reflected flux with phase angle, given by:

\begin{equation}
\psi(\alpha) = \frac{2}{3 \pi} a_s (\sin{\alpha} + \pi \cos{\alpha} - \alpha \cos{\alpha})
\end{equation} 

\noindent where $a_s$ is the surface albedo and $\alpha$ is the phase angle. The results are shown in figure
\ref{fig_phasefunc} where the integration has been carried out for grids of pixels with 30, 40 or 60 pixels across the
equator of the planet. The phase functions obtained by the integration method are almost identical to the analytic
version as shown in the upper panel of the plot. The small differences are shown on an expanded scale in the lower 
panel of the plot. The integrated results generally lie slightly below the analytic results with a little more structure
apparent at large phase angles where the number of pixels available becomes smaller at crescent phases.	The deviation
from the analytic function reaches a little over 1\% of the full phase value in the worst case.

\begin{figure*}
\includegraphics{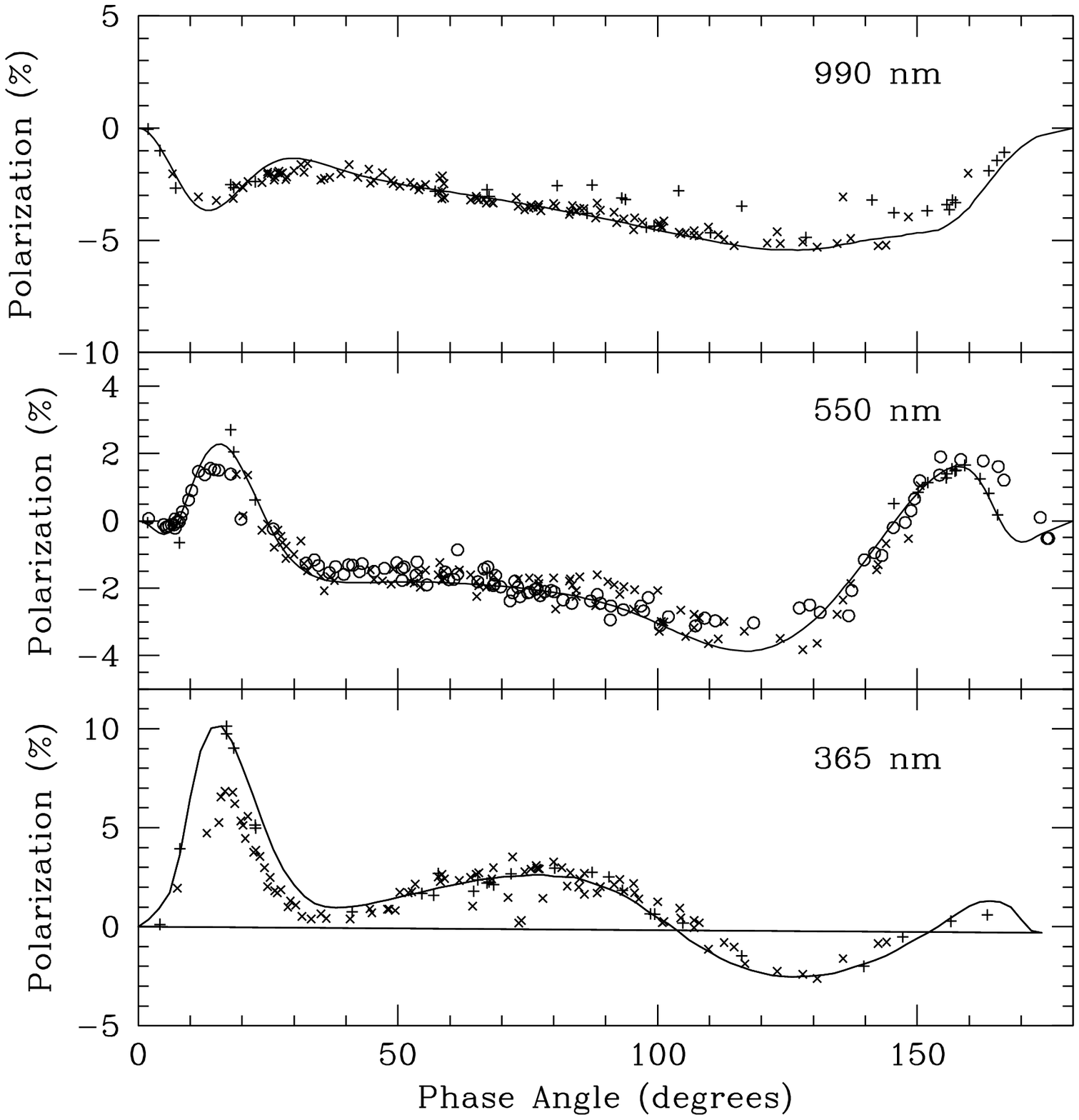}
\caption{\textsc{vstar/vlidort} models of the polarization phase curve of Venus compared with observations by 
\citet{lyot29} (circles), \citet{coffeen69} (crosses) and \citet{dollfus70} (plus signs). These models
reproduce the modelling of \citet{hansen74a} almost exactly and can be compared with figures 7 (990 nm),
4 (550 nm) and 9 (365 nm) of that paper. The models are for a single thick cloud layer containing
a gamma distribution of spherical particles with effective radius 1.05 $\mu$m and effective variance 0.07
 with refractive index and single scattering albedo as given in table \ref{tab_venus_clouds}. The
cloud layer also contains Rayleigh scatterers with the Rayleigh scattering optical depth being a fraction
$f_R = 0.045$ of the cloud optical depth at 365 nm.}
\label{fig_venus}
\end{figure*}

\section{The Polarization of Venus}

As a test of integrating polarization across the disc of a planet we have used our code to reproduce
the \citet{hansen74a} (hereafter HH74) study of the polarization phase curves of Venus. Venus is the
only planetary atmosphere in the Solar System that can be observed through its full range of phases
from the Earth. \citet{hansen71a} and HH74 modelled the polarization of Venus and fitted observations 
over a range of wavelengths. They found that the data could only be fitted if the cloud particles were
micron-sized liquid droplets with a refractive index consistent with a 75\% concentration of sulfuric
acid in water. Together with infrared reflectance measurements these results provided a strong case
for sulfuric acid being the main cloud constituent \citep{sill72,young73,pollack74}. These results
have been essentially confirmed by in-situ spacecraft observations.

\begin{table}
\caption{Venus cloud properties used in figure \ref{fig_venus}}
\label{tab_venus_clouds}
\begin{tabular}{lll}
\hline
Wavelength (nm) & $n_r$ & $\varpi$ \\
\hline
990   &   1.43   &   0.99941 \\
550   &   1.44   &   0.99897 \\
365   &   1.45   &   0.98427 \\
\hline
\end{tabular}
\end{table}

\begin{figure*} 
\includegraphics[angle=270,width=18cm]{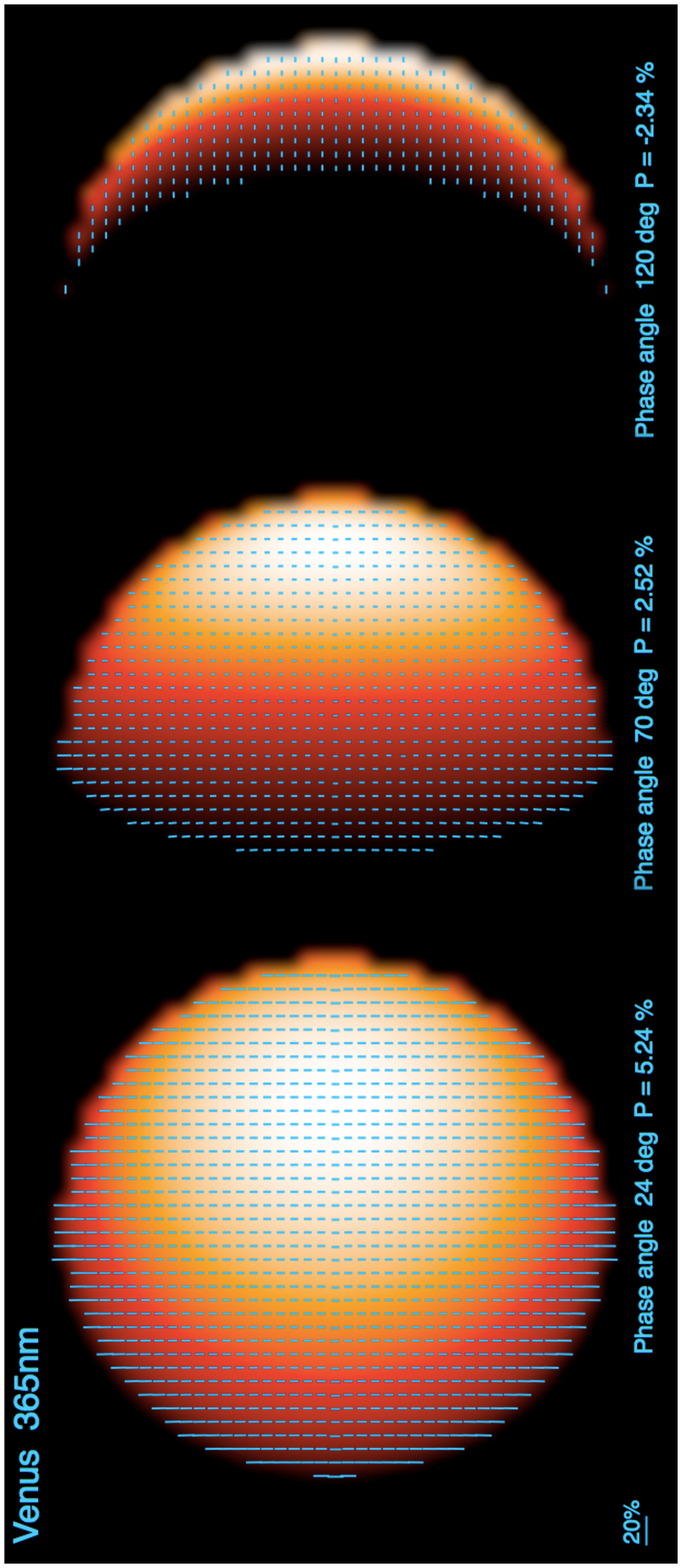} 
\caption{Images of the Venus
polarization models for selected phase angles from the same model data as used in the lower
panel of figure \ref{fig_venus}. Polarization vectors are shown overlaid on the intensity
distribution. The leftmost image is in the primary rainbow where the polarization is high, the
central image is in a region dominated by Rayleigh scattering, and the rightmost image shows a
phase where the polarization direction has reversed.} \label{fig_venus_im} \end{figure*}

As the main purpose of this test is as a verification of our code, we have reproduced the analysis of HH74
as closely as possible. They used a single layer model to represent the Venus atmosphere.
The clouds were modelled as a gamma
distribution \citep{hansen74, mishchenko02} of spherical  particles with effective radius
$r_{\mbox{\small eff}}$ = 1.05 $\mu$m, and effective variance $v_{\mbox{\small eff}}$ = 0.07. The real
refractive index ($n_r$) values are as given in table \ref{tab_venus_clouds}. The imaginary refractive
index was zero. However, instead of using the single scattering albedo ($\varpi$) derived from
Lorenz-Mie theory (this would be 1 for non-absorbing particles) the values of $\varpi$ in table
\ref{tab_venus_clouds} were used. HH74 derived these values as being those that made the geometric
albedo of the planet equal to the observed value. These make the cloud particles significantly
absorbing (particularly at 365 nm), and have the effect of increasing the polarization, since light
that is not single scattered, is more likely to  be absorbed, rather than returned via multiple
scattering with little polarization. This is, in effect, incorporating the UV absorber, known to be
present in the Venus atmosphere, into the same clouds responsible for the polarization. 

The total cloud optical depth was set to $\tau$ = 256 (as used by HH74 and described as equivalent to
$\infty$). As well as the clouds the layer includes Rayleigh scatterers with the Rayleigh scattering
optical depth set to a fraction $f_R$ = 0.045 of the cloud optical depth at 365 nm. In our model we used particles
with size much smaller than the wavelength to generate the Rayleigh scattering, although scattering from gas
molecules would have the same effect. 

Figure \ref{fig_venus} shows the resulting polarization phase curves integrated over the illuminated disc
compared with observations (mostly full disc polarization measurements) of the polarization of Venus.
Positive values refer to polarization perpendicular to the scattering plane, and negative values are
parallel to the scattering plane. The model phase curves are essentially identical to those of HH74. They
fit the observations reasonably well, and where they deviate a little from observations (e.g. the line is
a little below the data for 550 nm at phase angles $\sim$40 -- 70) they do so in the same way as the
models of HH74. Figure \ref{fig_venus_im} shows images of the polarization vectors overlaid on the intensity
image.

\begin{figure}
\includegraphics[width=\columnwidth]{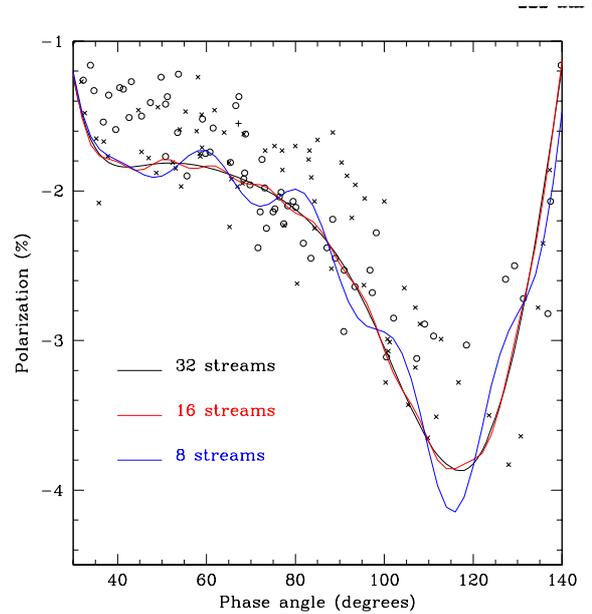}
\caption{Expanded view of part of the central panel (550 nm) from figure \ref{fig_venus} illustrating the ``ripple
pattern'' artifact in the phase curve that occurs when the discrete ordinate radiative transfer is run with too few
streams.}
\label{fig_ripple}
\end{figure}

It is important to run the calculations with sufficient streams. We used 48 streams in each
hemisphere for the 365 nm model, and 32 streams for the 550 and 990 nm. Running the model with too few
streams results in a ripple pattern artifact in the resulting phase curve. Figure \ref{fig_ripple}
shows the effect of reducing the number of streams to 16 or 8 for the 550 nm model. In this case the size parameter $x$
is $\sim$12 at 550 nm and $\sim$18 at 365 nm.

The agreement of our results with those of HH74 despite the use of different radiative transfer methods 
(HH74 used the doubling method) and the agreement of both with the observations provides a good test of our
approach.

\section{Exoplanet Polarization}

Here we consider the polarization of the transiting hot Jupiter HD 189733b. We choose this object because
it is one of the best studied exoplanets with a range of data that can be used to constrain an
atmospheric model. It also has evidence for the presence of scattering clouds both from transit spectroscopy
\citep{pont13}, and from a reflected light detection \citep{evans13}.

\subsection{HD 189733b Model}

Before we look at calculating the polarization of the reflected light, we first construct a
model of the atmosphere that is consistent with available data from the Hubble Space Telescope
(HST) and Spitzer Observatory on the secondary eclipse depths (emission spectrum) and transit
depths (transmission spectrum). In our models we assume solar abundances
\citep[from][]{grevesse07} and equilibrium
chemistry. We include absorption from molecules of H$_2$O, CO, CH$_4$, CaH, MgH, FeH and CrH,
atomic absorption from Na, K, Rb and Cs and collision induced absorption from H$_2$---H$_2$ and
H$_2$---He. We also include Rayleigh scattering from H, He and H$_2$, The methods and data
sources are as described in \citet{bailey12}, except for H$_2$---H$_2$ collision induced
absorption for which we have used the new HITRAN data \citep{richard12} based on \citet{abel11}.

\begin{figure}
\includegraphics[width=\columnwidth]{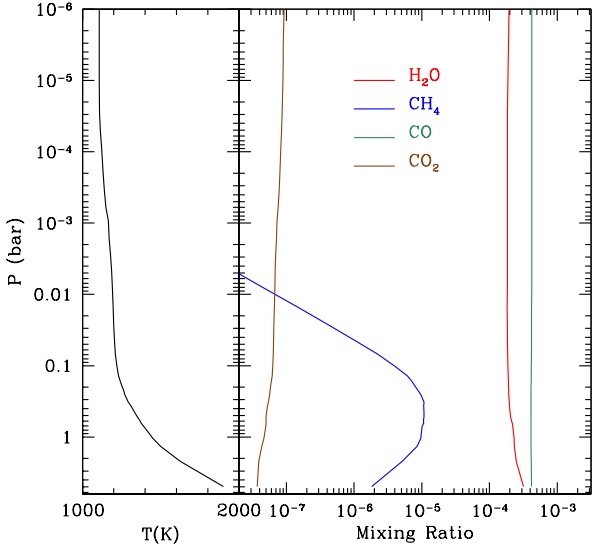}
\caption{Temperature profile and mixing ratios of important species for our model of HD 189733b}
\label{fig_model}
\end{figure}

We have adopted a temperature-pressure profile for the atmosphere similar to those found in
retrieval studies \citep[e.g.][]{lee12,line14}. However, we found we needed somewhat lower
temperatures in the 1 bar region than the retrieval models in order to fit the secondary 
eclipse data in the near-infrared region. The lower temperatures are a consequence of the lower
abundances for molecular species that result from our assumptions. Figure \ref{fig_model} shows
the adopted temperature profile and the mixing ratios for the species CO, H$_2$O, CH$_4$, and
CO$_2$. We note that equilibrium chemistry predicts that CO and H$_2$O are the most abundant
molecular species with relatively low abundances for CO$_2$ and CH$_4$. This is in agreement
with the results of high-resolution spectroscopy of HD 189733b, which detects CO and H$_2$O but
not CO$_2$ or CH$_4$, in both the dayside spectrum \citep{dekok13,birkby13} and the transmission
spectrum \citep{brogi16}.

\begin{figure}
\includegraphics[width = \columnwidth]{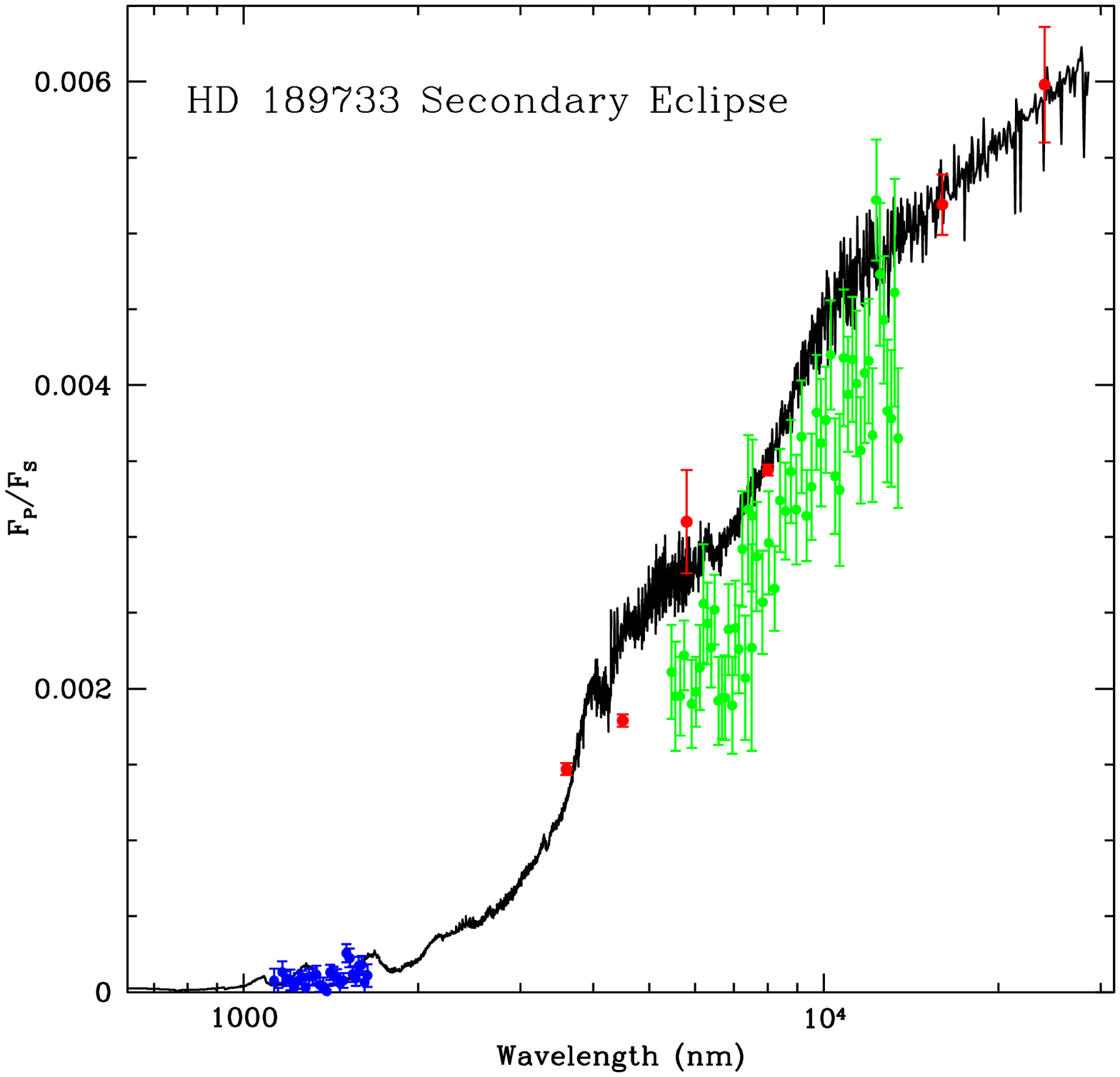}
\caption{Modelled emission spectrum of HD 189733b (expressed as planet flux divided by stellar flux,
F$_P$/F$_S$) compared with observations of the HD 189733b secondary eclipse. The observations
are, in red, Spitzer photometry from \citet{charbonneau08}, \citet{agol10} and
\citet{knutson12}, in green, Spitzer IRS spectroscopy from \citet{todorov14} and, in blue, HST
WFC3 spectroscopy from \citet{crouzet14}. The model also includes the same cloud and haze profile as
in figure \ref{fig_transit}.}
\label{fig_emission}
\end{figure}

In figure \ref{fig_emission} we show the emission spectrum predicted by this model compared with
observations of the secondary eclipses of HD 189733b. The spectrum is plotted as a fraction of
the stellar spectrum, using the Kurucz model\footnote{http://kurucz.harvard.edu/stars/hd189733/} 
for the spectrum of HD 189733 and a value
of 0.00033 for $(R_P/a)^2$ where $R_P$ is the planet radius and $a$ is the semi-major axis of the
orbit. The model spectrum is a good representation of the WFC3 data (blue points) and the
Spitzer photometry (red points). The latest analysis of the Spitzer IRS spectroscopy
\citep[][green points]{todorov14} lies a little below our model, but we prefer to match to the
Spitzer photometry that has much smaller errors.

\begin{figure}
\includegraphics[width = \columnwidth]{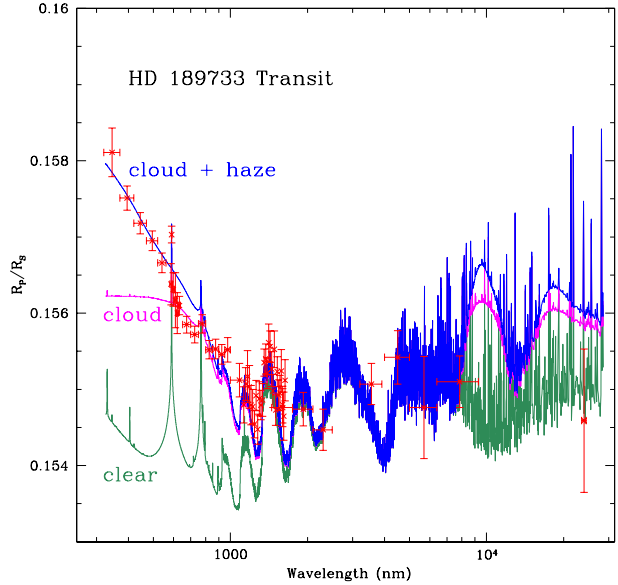}
\caption{Modelled transmission spectra of HD 189733b (expressed as planet radius divided by
stellar radius,
R$_P$/R$_S$) compared with observations of the HD 189733b transits. The observations, in red,
are Spitzer and HST data from \citet{pont13} and additional HST
WFC3 spectroscopy from \citet{mccullough14}. The models are for a cloud-free (clear) atmosphere,
a model with mid-level cloud only, and one with the same cloud plus a high-altitude haze. The
cloud and haze here are modelled as 0.05 $\mu$m enstatite particles.}
\label{fig_transit}
\end{figure}

The clouds in the model are adjusted to fit to the observed transit data. We find that the clouds at the required
level have very little effect on the emission spectrum which is primarily determined by the temperature profile.
Similarly changes to the temperature profile have only a small effect on the transmission spectrum (mostly through
the dependence of the scale height on temperature). Thus it requires only a few iterations to derive a model that
fits both emission and transmission spectra.

We use transit data from \citet{pont13} as well as additional HST/WFC3
data from \citet{mccullough14}. To fit the transmission data we need both a haze component at
high altitudes (10$^{-2}$ to 10$^{-4}$ mbar) as well as a thicker cloud layer at a lower altitude
of 0.35 to 3.5 mbar. Figure \ref{fig_transit} shows that a clear model does not fit the data at
all well over the visible wavelengths. There is a small rise to the blue caused by Rayleigh
scattering from molecules, but overall the clear model falls well below the observations from 
0.3 to about 1.4 $\mu$m. Adding the cloud layer improves the fit from about 0.7 to 1.4 $\mu$m,
but the high altitude haze is needed to also fit the data points in the blue.

We note that \citet{mccullough14} have proposed an alternate expalanation of the
transmission spectrum of HD 189733b, in which unocculted starspots, are responsible for the steep
rise at blue wavelengths allowing a clear atmosphere to be consistent with the data.
\citet{sing16} have looked at the transmission spectra of a larger sample of hot Jupiters and find
no correlation between the slope of the transmission feature in the blue, and the stellar
activity, thus favouring the cloud interpretation. 

We have tried various different compositions for the cloud particles. These include pure
Rayleigh particles (i.e. very small, non-absorbing particles) as well as several condensate species
that are expected, from chemical models, to be important in hot Jupiter atmospheres. These are
the silicates enstatite (MgSiO$_3$) and forsterite (Mg$_2$SiO$_4$) with refractive indices from
\citet{scott96}, corundum \citep[Al$_2$O$_3$, refractive indices from][]{koike95} and iron
\citep[Fe, refractive indices from][]{johnson74,ordal88}. For these species we used a power law
size distribution of particles as defined by \citet{hansen74} with effective variance $v_{\mbox{\small eff}}$ = 0.01
and a range of effective radii.

\begin{figure}
\includegraphics[width = \columnwidth]{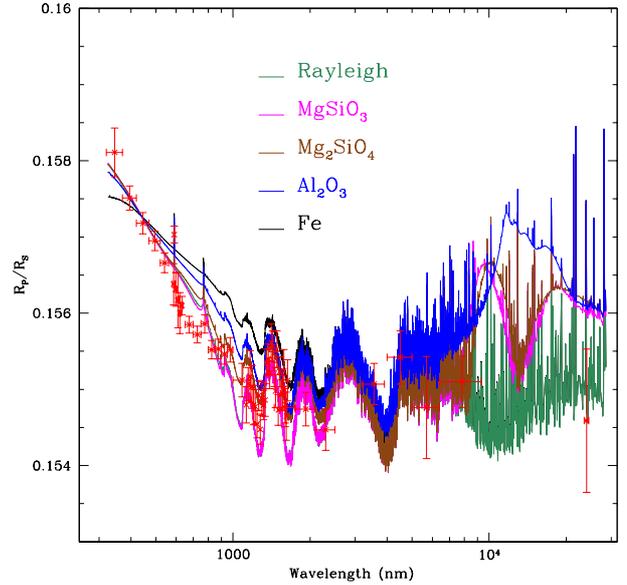}
\caption{As figure \ref{fig_transit}, but showing models with different particle compositions,
either pure Rayleigh particles (green) or particles with $r_{\mbox{\small eff}}$ = 0.05 $\mu$m composed of
enstatite (MgSiO$_3$), forsterite (Mg$_2$SiO$_4$), corundum (Al$_2$O$_3$) and iron (Fe). In all
cases the same optical depth profile for the cloud at 440 nm as that in figure \ref{fig_transit}
 has been used. }
\label{fig_composition}
\end{figure}

Figure \ref{fig_composition} shows a comparison of the predicted transmission spectrum for the
different particle compositions. An effective radius of $r_{\mbox{\small eff}}$ = 0.05 $\mu$m has been used in all cases,
and the cloud and haze optical depth profile is the same as used in figure \ref{fig_transit} and
specified at 440 nm wavelength. Note that all the particle types produce a rise in the blue similar
to that observed, and with some adjustment of the optical depth profile could probably be made
to match the observations better.

It can also be seen from figure \ref{fig_composition} that there are substantial differences in the
transmission spectrum in the region from 6 to 20 $\mu$m where distinct absorption features are
present. This region is not currently well constrained by observations, but future observations
with the James Webb Space Telescope (JWST) at these wavelengths could help to determine cloud particle
compositions and sizes \citep{wakeford15}.

The model we have obtained here is not unique, and clearly contains some simplifications. In
particular we have assumed that the same model structure applies to the dayside and to the
terminator region probed by transmission spectroscopy, when it is known that HD 189733b has
temperature structure seen in infrared phase curves \citep{knutson07,ddixon13}. However, the
main point here is to show how we can use the same code to model the emission spectrum,
transmission spectrum, and reflected light phase curves and polarization phase curves, and to
calculate the polarization using a fully realistic model which is consistent with other data.

\subsection{Polarization Phase Curves}

\begin{figure}
\includegraphics[width = \columnwidth]{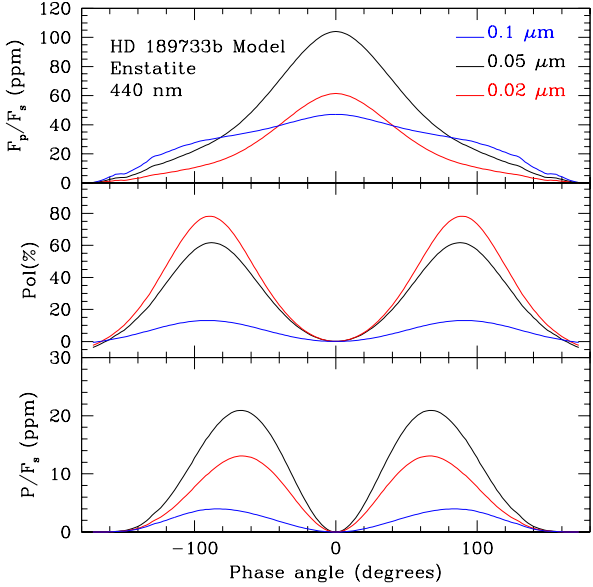}
\caption{Phase curves and polarization for HD 189733b models with the same cloud/haze optical depth
profile as used in figures \ref{fig_transit} and \ref{fig_composition} for enstatite
particles of various sizes. Top panel is the reflected flux from the planet as a fraction of that from
the star. Middle panel is the percentage polarization of the integrated reflected light. Lower panel
is the polarization fraction in ppm that would be observed in the combined light of the planet and
star.}
\label{fig_phase_enst}
\end{figure}

\begin{figure}
\includegraphics[width = \columnwidth]{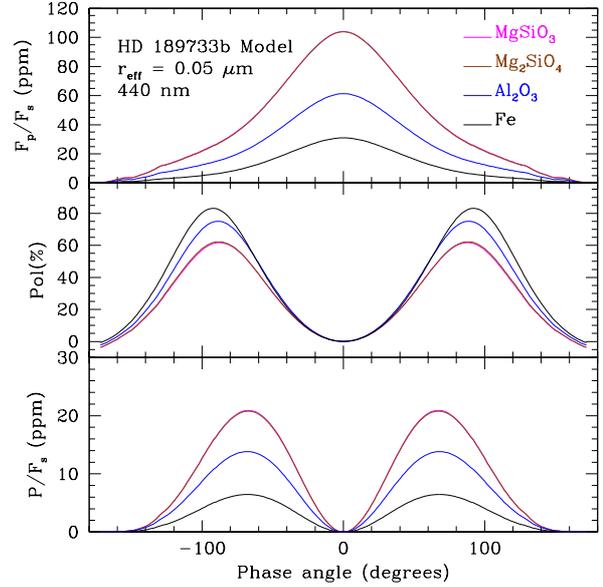}
\caption{As figure \ref{fig_phase_enst} but showing the effect of different haze/cloud particle
compositions all with r$_{\mbox{\small eff}}$ = 0.05 $\mu$m. The curves for the sliciates enstatite and
forsterite are almost identical.}
\label{fig_phase_comp}
\end{figure}

We can now calculate the polarization and reflected light phase curves for our model of HD 189733b
using the methods outlined in sections 2 and 4. As we are dealing with quite small cloud particles
close to the Rayleigh regime, a relatively small number of streams is sufficient. We ran these models
with 8 stream in each hemisphere. Figure \ref{fig_phase_enst} shows the phase curves for our model using the
same cloud/haze optical depth profile as used in figures \ref{fig_transit} and \ref{fig_composition}.
The calculations are for 440 nm, a wavelength typically used for observations of the polarization of HD
189733. In this and subsequent plots the top panel shows the reflected light phase curve in
parts-per-million (ppm), the middle panel shows the integrated  polarization of the planet's light in
per cent, and the bottom panel shows the polarization as a fraction of the light from the star. This
is what we would actually observe in a measurement of the polarization of the combined light of the
planet and star, if the star is assumed unpolarized. We will refer to this as ``observed
polarization'' in subsequent discussions.

The peak value shown in the top panel, which occurs at zero phase angle, divided by $(R_P/a)^2$ (0.00033 in our
models of HD 189733b) gives the geometric albedo of the planet. For example, a value of 100 ppm (or 0.0001)
corresponds to a geometric albedo of 0.303.

Figure \ref{fig_phase_enst} shows the effect of particle size for enstatite particles. Particles of 0.05 $\mu$m effective
radius produce the highest amplitude for both the phase curve and observed polarization. This is
becuase this particle size provides the closest to Rayleigh-like behaviour at this wavelength.
Increasing the size to 0.1 $\mu$m moves the particles into the Mie-scattering regime resulting in
substantially reduced amplitudes and different phase behaviour. Particles of 0.02 $\mu$m or smaller
become increasingly absorbing and this reduces light curve and polarization curve amplitude, although
the percentage polarization of the reflected light increases.

Figure \ref{fig_phase_comp} shows the effect of particle compositions for 0.05 $\mu$m radius. The
silicate particles, enstatite and forsterite, produce almost identical curves. Corundum and Iron
particles produce lower amplitudes in reflected light and observed polarization.

\begin{figure}
\includegraphics[width = \columnwidth]{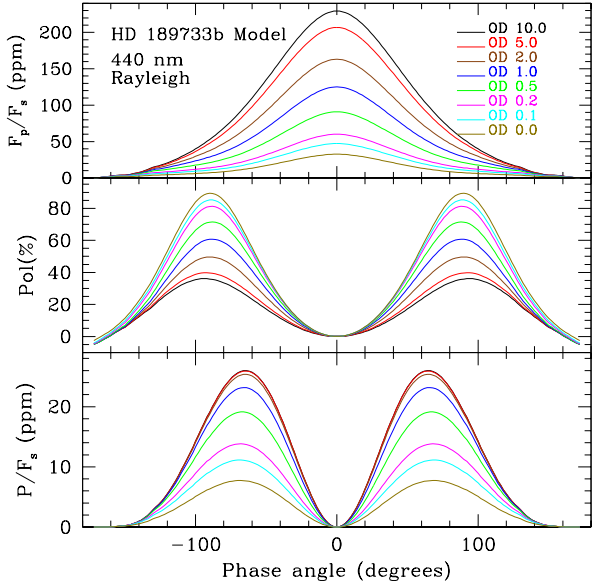}
\caption{Phase curves and polarization for HD 189733b models with a single cloud layer of pure
Rayleigh particles with cloud optical depth varied
from 0.0 to 10.0. The observed polarization amplitude (lower panel) is highest for the thickest cloud.
Decreasing the cloud optical depth increases the fractional polarization of the reflected light, but
decreases the albedo of the planet and so decreases the observed polarization.}
\label{fig_od_ray}
\end{figure}

\begin{figure}
\includegraphics[width = \columnwidth]{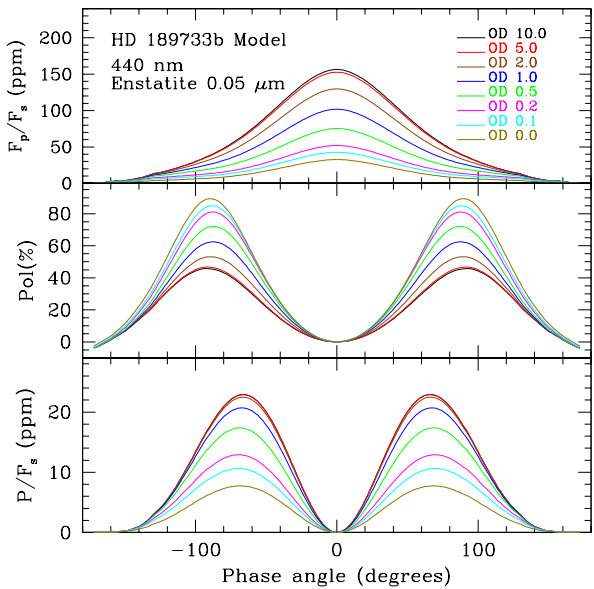}
\caption{As figure \ref{fig_od_ray} but for clouds composed of 0.05 $\mu$m enstatite particles. The
enstatite particles behave similarly to the pure Rayleigh case, but with slightly lower albedo
and observed polarization. }
\label{fig_od_enst05}
\end{figure}

In figures \ref{fig_od_ray} and \ref{fig_od_enst05} we show the effect of changing the optical depth
of the cloud (at 440 nm) from $\tau$ = 0 to 10 for Rayleigh particles and enstatite particles of 0.05
$\mu$m effective radius. In this case we place the cloud in a single layer. It can be seen that the
light curve and polarization curve amplitudes are largest for optically thick clouds. As the optical
depth of the cloud decreases the percentage polarization of the planetary reflected light increases
but the light curve amplitude and consequently observed polarization decreases.

\subsection{Polarization Wavelength Dependence}

\begin{figure}
\includegraphics[width = \columnwidth]{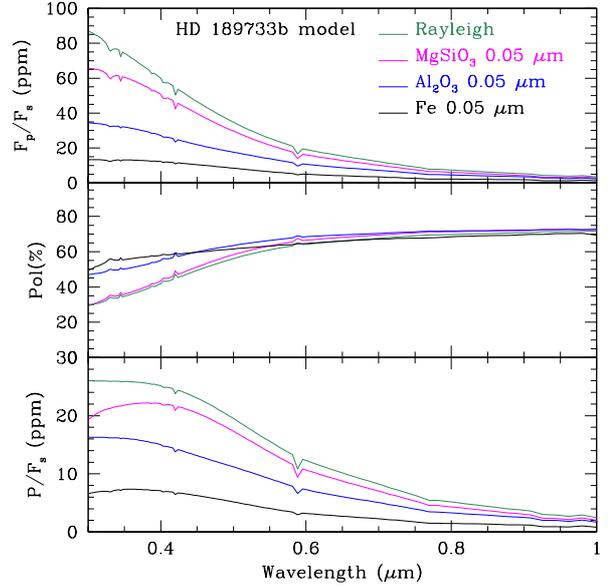}
\caption{Polarization wavelength dependence for HD 189733b models with the same cloud/haze optical 
depth profile as used in figures \ref{fig_transit} and \ref{fig_composition}. The results are plotted
for phase angle 67 degrees which is approximately where the observed polarization normally peaks. Only
the contribution of reflected light is shown here. There will be significant contributions also from
thermal emission from the planet at the red end of this plot.}
\label{fig_pol_spec_comp}
\end{figure}

Figure \ref{fig_pol_spec_comp} shows the wavelength dependence of the reflected light from the planet
and the polarization as seen at a phase angle of 67 degrees, where the observed polarization reaches
its peak. The plots are for the cloud optical depth profile used in figures \ref{fig_transit} and 
\ref{fig_composition}. It can be seen that in general the reflected light fraction and polarization
increases to the blue. However, for the enstatite case there is a drop in polarization at the shortest
wavelengths as the particles become more absorbing.

\subsection{Discussion}

The results of the models presented above can be understood as follows. Light that is single scattered
from the cloud particles will generally be the most highly polarized. In an optically thick Rayleigh
scattering cloud, light that is not single scattered will undergo multiple scattering in the cloud and
will eventually be reflected outward after multiple scattering events. This light will tend to have low
polarization as the random orientations of the multiple scattering tend to cancel polarization.
Models by \citet{buenzli09} showed that an optically thick Rayleigh scattering atmosphere produces 
a maximum polarization of
32.6 per cent due to this dilution by multiply-scattered light.

Reducing the optical depth of the cloud increases the chance that light that fails to single scatter
upwards will pass through the cloud and be absorbed in lower layers of the atmosphere. This increases
the polarization of the reflected light, becuase there is a higher contribution from single scattered
light. However, as more light is absorbed, the albedo of the planet and the observed polarization will
decrease as a result of this effect.

Making the cloud particles themselves more absorbing, or mixing them with absorbing gases, has the
same effect. Light that fails to single scatter is more likely to be absorbed before it can multiple
scatter upwards. Once again this increases the polarization of the reflected light but decreases the
amount of reflected light and hence the observed polarization.

\subsection{Comparison with Previous Modelling of HD 189733b}

A key result of our modelling is that the maximum polarization amplitude for HD 189733b is about 27 ppm.
An early model by \citet{sengupta08} that considered only single scattering gave a polarization amplitude of more 
than 100 ppm. However, as we show in figure \ref{fig_od_ray}  the onset of multiple scattering limits the polarization to
much lower values. \citet{lucas09} used a Monte-Carlo radiative transfer approach to obtain an upper limit on the
polarization of HD 189733b of 26 ppm, in good agreement with our results. \citet{berdyugina11} considered a simple model
in which the planet reflected as a Lambert sphere with a geometric albedo of 2/3 and was maximally polarized (i.e. 100\%)
in order to explain the large observed polarization described below. However, as can be seen from figure \ref{fig_od_ray}
this combination is not possible. A large geometric albedo can be obtained with optically thick clouds, and a high
polarization can be obtained with optically thin clouds, but the two cannot be obtained together.

\citet{kopparla16} present a multiple scattering radiative transfer model for HD 189733b, which, like ours, is based
on the \textsc{vlidort} code. Their reported polarization amplitudes range from about 40 to 60 ppm, quite different to
ours, and above our limit of 27 ppm. However, on examining their results we find that their modelled fluxes for the planet
at phase zero are inconsistent with the values expected for the geometric albedo they assume. The flux of the planet as a
fraction of that from the star ($f$) is related to the geometric albedo $A_g$ by:

\begin{equation}
f = (R_p/a)^2 A_g
\end{equation}

For HD 189733b $(R_p/a)^2 = 0.00033$ so for the geometric albedo of 0.23 which \citet{kopparla16} adopt for normalization
of their models $f$ should be 76 ppm. In fact the zero phase value shown for their models (e.g. their figure 4) is about
200 ppm. It appears that a scaling error has occurred in the normalization of their model results and all their ppm figures
for intensity and polarization are a factor of about 2.6 too high. When their results are corrected for this factor they
agree quite well with our results.

\subsection{Comparison with Observations}

\citet{evans13} have reported a reflected light detection of HD 189733b with HST through an observed
secondary eclipse depth of 126$^{+37}_{-36}$ ppm for a wavelength of 290--450 nm. This can be compared
with the peak value at zero phase angle in the top panel of figures \ref{fig_phase_enst} --
\ref{fig_od_enst05}. It agrees quite well with our value for 0.05 $\mu$m particles and cloud optical
depth around 1. Any of our models that matches a 126 ppm reflected light signal, has an observed polarization
of $\sim$20 ppm or a little higher. While the wavelength of the HST observation is a little shorter than the 440
nm used for most of our models, we note that figure \ref{fig_pol_spec_comp} does not show much change in polarization
with wavelength over this range. 

A planetary polarization signal from HD 189733b with an amplitude of $\sim$100 ppm has been reported by
\citet{berdyugina11}. However, a signal of this amplitude is not seen in other polarization observations
of this object 
\citep{wiktorowicz09,wiktorowicz15,bott16}. Based on the modelling here we find that such a signal is
too large by a factor of four to be explained as due to Rayleigh scattering from the atmosphere of HD
189733b.

Our modelling suggests a more plausible amplitude for the polarization of HD 189733b at 440 nm is
$\sim$20 ppm. This is entirely consistent with the observations of \citet{wiktorowicz15} and
\citet{bott16} and with the tentative signal of 29.4 $\pm$ 15.6 ppm reported in the latter work.
Further observations are needed to provide a definitive test of the presence of polarized reflected
light from this planet.

There are, however, a variety of ways in which the polarization could be lower. If there are no clouds, as in the
\citet{mccullough14} interpretation of the transit data, the polarization drops to only a few ppm (due
to Rayleigh scattering from H, He and H$_2$). The
polarization can also be low as a result of the cloud particles being too large (see figure \ref{fig_phase_enst})
or being more absorbing (figure \ref{fig_phase_comp}). All these cases would also result in the reflected light
intensity being reduced to well below the level reported by \citet{evans13}. However, the uncertainties on that
measurement are such that this cannot be excluded.

\begin{figure}
\includegraphics[width = \columnwidth]{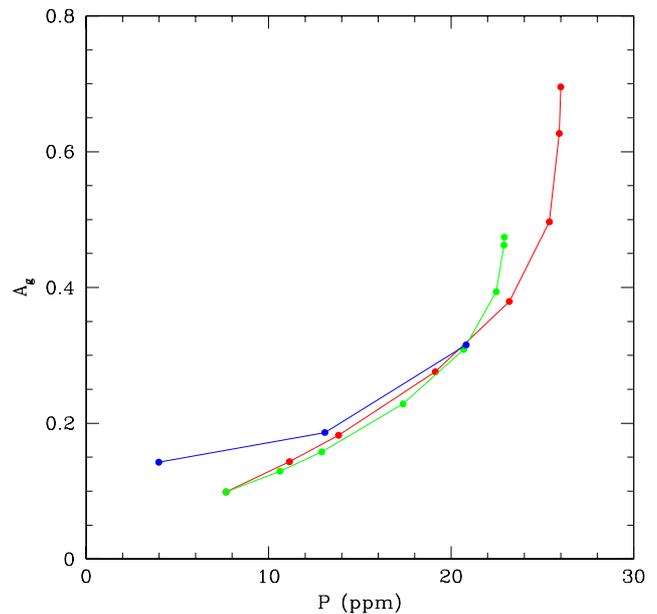}
\caption{Geometric albedo of planet plotted against polarization amplitude for several of the models of HD 189733b considered
previously. Red - models of pure Rayleigh scattering particles as in figure \ref{fig_od_ray}. Green - models for 0.05
$\mu$m enstatite particles as in figure \ref{fig_od_enst05}. Blue - models for different sized enstatite particles as in
figure \ref{fig_phase_enst}.}
\label{fig_hd189_albedo}
\end{figure}

Another complication with any attempt to detect polarization from HD 189733b is that HD 189733 is an active BY Dra
type K dwarf. Recent work by \citet{cotton17b} shows that active K dwarfs are typically polarized at levels of
$\sim$ 30 ppm, while inactive dwarfs have low polarization. The polarization is most likely the result of
differential saturation of spectral lines in the presence of magnetic fields. The stellar polarization is likely
to be variable and will confuse any attempt to pick out the planetary signal.

Thus more favourable targets for detecting exoplanet polarization might be hot Jupiters orbiting inactive stars
that can be expected to have low polarizations \citep{cotton17b}. The polarization technique is not limited to
transiting planets. The results in figures \ref{fig_phase_enst} to \ref{fig_od_enst05} can be used as a guide to
estimate the polarization of other planets by scaling the top and bottom panels in proportion to $(R_P/a)^2$,
which has value 0.00033 for the HD 189733b models shown.

\subsection{Polarization and Geometric Albedo}

A number of previous studies have attempted to use limits on the observed polarization of an exoplanet to set limits on
the geometric albedo of the planet \citep{lucas09,wiktorowicz09,wiktorowicz15}. In figure \ref{fig_hd189_albedo} we show
the relationship between geometric albedo $A_g$ and polarization amplitude for a number of the models we have considered
in this study. It can be seen that though there is a general trend for high polarization amplitudes to correspond to high
geometric albedos the precise relationships are dependent on the assumptions about particle size and composition.

Thus in general it is hard to set definitive limits on geometric albedo based on polarization observations without
making assumptions about the particle properties. In particular most such comparisons are only valid for small
Rayleigh-like particles. The example of Venus considered earlier, shows how a high geometric albedo can be achieved with
realtively low polarization amplitudes, although the Venus cloud properties are very different from those we expect to
find in hot Jupiters.

We further note that the specific limit on geometric albedo of HD 189733b of $A_g$ < 0.40 reported by
\citet{wiktorowicz15} is based on the same modelling as that of \citet{kopparla16} which contains a scaling error as
noted earlier. When this scaling error is corrected no siginifcant upper limit on the geometric albedo can be set based
on these observations.

We can reverse the analysis and used the observed geometric albedo obtained by \citet{evans13} together with
figure \ref{fig_hd189_albedo} to obtain aother estimate of the predicted polarization amplitude of HD 189733b of $\sim$22
ppm, in reasonable agreement with our earlier estimate. As noted above this estimate is based on the assumption of
small Rayleigh-like cloud particles.

\section{Conclusions}

We have described the modifications required to our \textsc{vstar} radiative transfer code to incorporate
polarized radiative transfer. The resulting code enables us to predict the disc-resolved, phase-resolved and
spectrally-resolved intensity and polarization for any of the wide range of planetary, substellar and cool-star
atmospheres that can be modelled with \textsc{vstar}. The polarized radiative transfer equation is solved using
the \textsc{vlidort} code of \citet{spurr06} that uses the discrete-ordinate method.

We have tested the code by using it to reproduce a standard benchmark problem in polarized radiative transfer
\citet{garcia89} and find agreement with the benchmark results to 5-6 significant figures. We have also reproduced
the calcuations of the polarization phase curves of Venus carried out by \citet{hansen74a} and obtain essentially
identical results.

We have used the code to model the polarization of the hot Jupiter HD 189733b. We first construct a model of the
atmosphere that is consistent with the observed emission and transmission spectra. We then calculate the
polarization phase curves predicted by that model. We are able to use the same code to model the emission,
transmission, reflection and polarization properties of the atmosphere. We predict a polarization amplitude of
$\sim$20 ppm for a model consistent with other observations including the \citet{evans13} detection of reflected
light. The maximum polarization amplitude we can obtain is $\sim$27 ppm achieved with optically thick Rayleigh
scattering clouds. The predictions are consistent with polarization observations reported by \citet{bott16} and
\citet{wiktorowicz15}.

The detection of polarization in HD189733 at about the predicted level would be strong evidence for the presence of clouds and
would rule out clear models such as that of \citet{mccullough14}. A clear atmosphere would result in a low polarization
amplitude of only a few ppm. Intermediate levels of polarization
may indicate that the cloud particles are smaller or more absorbing. In this case
the reflected light intensity will drop below the value reported by \citet{evans13}. The results can be scaled to
other hot Jupiter systems. We note that hot Jupiters orbiting inactive stars may be better targets for
polarimetric detection than HD 189733b due to the lower polarization of the host star \citep{cotton17b}.

\section*{Acknowledgements}

This work was supported by the Australian Research Council through Discovery Project grants
DP110103167 and DP160103231. We thank Robert Spurr of RT Solutions for making available the \textsc{vlidort}
software. We thank Daniel Cotton for comments on the manuscipt, and an anonymous referee for suggestions that
significantly improved the paper.





\begin{thebibliography}{99}
\bibitem[\protect\citeauthoryear{Abel et al.}{2011}]{abel11} Abel, M., Frommhold, L., Li, X.,
Hunt, K.L.C., 2011, J. Phys. Chem. A, 115, 6805
\bibitem[\protect\citeauthoryear{Agol et al.}{2010}]{agol10} Agol, E., Cowan, N.B., Knutson,
H.A., Deming, D., Steffen, J.H., Henry, G.W., Charbonneau, D., 2010, ApJ, 721, 1861
\bibitem[\protect\citeauthoryear{Bailey}{2007}]{bailey07} Bailey, J., 2007, Astrobiology, 7, 320
\bibitem[\protect\citeauthoryear{Bailey}{2009}]{bailey09} Bailey, J., 2009, Icarus, 201, 444
\bibitem[\protect\citeauthoryear{Bailey \& Kedziora-Chudczer}{2012}]{bailey12} Bailey, J., 
Kedziora-Chudczer, L., 2012, MNRAS, 419, 1913
\bibitem[\protect\citeauthoryear{Bailey, Simpson \& Crisp}{Bailey et al.}{2007a}]{bailey07a} Bailey,
J., Simpson, A., Crisp, D., 2007, PASP, 119, 228
\bibitem[\protect\citeauthoryear{Bailey et al.}{2008}]{bailey08} Bailey, J., Meadows, V.S.,
Chamberlain, S., Crisp, D., 2008b, Icarus, 197, 247
\bibitem[\protect\citeauthoryear{Bailey, Ahlsved \& Meadows}{Bailey et al.}{2011}]{bailey11} Bailey, J.,
Ahlsved, L., Meadows, V.S., 2011, Icarus, 213, 218
\bibitem[\protect\citeauthoryear{Bailey et al.}{2015}]{bailey15} Bailey, J., Kedziora-Chudczer,
L., Cotton, D.V., Bott, K., Hough, J.H., Lucas, P.W., 2015, MNRAS, 449, 3064
\bibitem[\protect\citeauthoryear{Berdyugina et al.}{2011}]{berdyugina11} Berdyugina, S.V., Berdyugin,
A.V., Fluri, D.M., Piirola, V., 2011, ApJL, 728, L6
\bibitem[\protect\citeauthoryear{Birkby et al.}{2013}]{birkby13} Birkby, J.L., de Kok, R.J.,
Brogi, M., de Mooij, E.J.W., Schwarz, H., Albrecht, S., Snellen, I.A.G., 2013, MNRAS, 436, L35
\bibitem[\protect\citeauthoryear{Brogi et al.}{2016}]{brogi16} Brogi, M., de Kok, R.J., Albrecht, S.,
Snellen, I.A.G., Birkby, J.L., Schwarz, H., 2016, ApJ, 817, 106
\bibitem[\protect\citeauthoryear{Bott et al.}{2016}]{bott16} Bott, K., Bailey, J., Kedziora-Chudczer,
L., Cotton, D.V., Lucas. P.W., Marshall, J.P., Hough, J.H., 2016, MNRAS, 459, L109 
\bibitem[\protect\citeauthoryear{Buenzli \& Schmid}{2009}]{buenzli09} Buenzli, E., Schmid,
H.M., 2009, A\&A, 504, 259
\bibitem[\protect\citeauthoryear{Chamberlain et al.}{2013}]{chamberlain13} Chamberlain, S., Bailey, J., Crisp, D.,
Meadows, V.S., 2013, Icarus, 222, 364
\bibitem[\protect\citeauthoryear{Charbonneau et al.}{2008}]{charbonneau08} Charbonneau, D.,
Knutson, H.A., Barman, T., Allen, L.E., Mayor, M., Megeath, S.T., Queloz, D., Udry, S., 2008,
ApJ, 686, 1341.
\bibitem[\protect\citeauthoryear{Coffeen \& Gehrels}{1969}]{coffeen69} Coffeen, D.L., Gehrels, T.,
1969, A\&A, 74, 433
\bibitem[\protect\citeauthoryear{Cotton et al.}{2012}]{cotton12} Cotton, D.V., Bailey, J., Crisp, D., Meadows, V.S.,
2012, Icarus, 217, 570
\bibitem[\protect\citeauthoryear{Cotton, Bailey, \& Kedziora-Chudczer}{2014}]{cotton14} Cotton, D.V., Bailey, J.,
Kedziora-Chudczer, L., 2014, MNRAS, 439, 387
\bibitem[\protect\citeauthoryear{Cotton et al.}{2017a}]{cotton17a} Cotton, D.V., Bailey, J., Howarth, I.D., Bott,
K., Kedziora-Chudczer, L., Lucas, P.W., Hough, J.H., 2017a, Nature Ast., 1, 690
\bibitem[\protect\citeauthoryear{Cotton et al.}{2017b}]{cotton17b} Cotton, D.V., Marshall, J.P., Bailey, J.,
Kedziora-Chudczer, L., Bott, K., Marsden, S.C., Carter, B.D., 2017b, MNRAS, 467, 873
\bibitem[\protect\citeauthoryear{Crouzet et al.}{2014}]{crouzet14} Crouzet, N., McCullough, P.R., Deming,
D., Madhusuhan, N., 2014, ApJ, 795, 166
\bibitem[\protect\citeauthoryear{Cuesta et al.}{2013}]{cuesta13} Cuesta, J., et al., Atmos. Chem. Phys., 13, 9675
\bibitem[\protect\citeauthoryear{de Kok et al.}{2013}]{dekok13} de Kok, R.J., Brogi, M.,
Snellen, I.A.G., Birkby, J., Albrecht, S., de Mooij, E.W.J., 2013, A\&A, 554, A82
\bibitem[\protect\citeauthoryear{de Rooij \& van der Stap}{1984}]{rooij84} de Rooij, W.A., van der
Stap, C.C.A.H., 1984, AJ, 131, 237
\bibitem[\protect\citeauthoryear{Dobbs-Dixon \& Agol}{2013}]{ddixon13} Dobbs-Dixon, I., Agol,
E., 2013, MNRAS, 435, 3159
\bibitem[\protect\citeauthoryear{Dollfus \& Coffeen}{1970}]{dollfus70} Dollfus, A., Coffeen, D.L.,
1970, A\&A, 8, 251
\bibitem[\protect\citeauthoryear{Evans \& Stephens}{1991}]{evans91}
Evans, K.F. \& Stephens, G.L., 1991, \jqsrt, 46, 415.
\bibitem[\protect\citeauthoryear{Evans et al.}{2013}]{evans13} Evans, T.M. et al., 2013, ApJ,
772, L16
\bibitem[\protect\citeauthoryear{Garcia \& Siewert}{1989}]{garcia89} Garcia, R.D.M., \& Siewert,
C.E., J. Quant. Spectrosc. Radiative Transfer, 41, 117
\bibitem[\protect\citeauthoryear{Grevesse, Asplund \& Sauval}{2007}]{grevesse07} Grevesse, N.,
Asplund, M., Sauval, A.J., 2007, SSRv, 130, 15
\bibitem[\protect\citeauthoryear{Hammer et al.}{2016}]{hammer16} Hammer, M.S., Martin, R.V., van Donkelaar, A.,
Buchard, V., Torres, O., Ridley, D.A., Spurr, R.J.D., 2016, Atmos. Chem. Phys., 16, 2507
\bibitem[\protect\citeauthoryear{Hansen \& Hovenier}{1971}]{hansen71} Hansen, J.E., \& Hovenier, J.W., 1971,
J. Quant. Spectrosc. Radiative Transfer, 11, 809
\bibitem[\protect\citeauthoryear{Hansen \& Arking}{1971}]{hansen71a} Hansen, J.E., \& Arking, A., 1971,
Science, 171, 669
\bibitem[\protect\citeauthoryear{Hansen \& Hovenier}{1974}]{hansen74a} Hansen, J.E. \& Hovenier, J.W., 1974, J.
Atmos. Sci., 31, 1137
\bibitem[\protect\citeauthoryear{Hansen \& Travis}{1974}]{hansen74} Hansen, J.E. \& Travis,
L.D., 1974, Space Sci. Rev., 16, 527
\bibitem[\protect\citeauthoryear{Harrington}{2015}]{harrington15} Harrington, J.P., 2015, Proc. IAU, 305, 395
\bibitem[\protect\citeauthoryear{Horak}{1950}]{horak50} Horak, H.G., 1950, ApJ, 112, 445
\bibitem[\protect\citeauthoryear{Hough et al.}{2006}]{hough06} Hough, J.H., Lucas, P.W., Bailey,
J.A., Tamura, M., Hirst, E., Harrison, D., Bartholomew-Briggs, M., 2006, PASP, 118, 1302
\bibitem[\protect\citeauthoryear{Johnson \& Christy}{1974}]{johnson74} Johnson, P.B., Christy,
R.W., 1974, \prb 9, 5056
\bibitem[\protect\citeauthoryear{Karalidi \& Stam}{2012}]{karalidi12} Karalidi, T. \& Stam,
D.M., 2012, A\&A, 546, A56
\bibitem[\protect\citeauthoryear{Karalidi, Stam \& Guirado}{Karalidi et al.}{2013}]{karalidi13}
 Karalidi, T., Stam, D.M., Guirado, D., 2013, A\&A, 555, A127
\bibitem[\protect\citeauthoryear{Kedziora-Chudczer \& Bailey}{2011}]{chudczer11}
Kedziora-Chudczer, L., and Bailey, J., 2011, MNRAS, 414, 1483
\bibitem[\protect\citeauthoryear{Knutson et al.}{2007}]{knutson07} Knutson, H.A. et al., 2007,
Nature, 447, 183
\bibitem[\protect\citeauthoryear{Knutson et al.}{2012}]{knutson12} Knutson, H.A. et al., 2012,
ApJ, 754, 22
\bibitem[\protect\citeauthoryear{Koike et al.}{1995}]{koike95} Koike, C., Kaito, C., Yamamoto,
T., Shibai, H., Kimura, S., Suto, ., 1995, Icarus, 114, 203
\bibitem[\protect\citeauthoryear{Kopparla et al.}{2016}]{kopparla16}Kopparla, P., Natraj, V., Zhang, X., Swain, M.R., Wiktorowicz,
S.J., Yung, Y.L., 2016, ApJ, 817, 32
\bibitem[\protect\citeauthoryear{Lee, Fletcher \& Irwin}{Lee et al.}{2012}]{lee12} Lee, J.-M.,
Fletcher, L.N., Irwin, P.G.J., 2012, MNRAS, 420, 170
\bibitem[\protect\citeauthoryear{Line et al.}{2014}]{line14} Line, M.R., Knutson, H., Wolf,
A.S., Yung, Y.L., 2014, ApJ, 783, 70
\bibitem[\protect\citeauthoryear{Lucas et al.}{2009}]{lucas09} Lucas, P.W., Hough, J.H., Bailey,
J.A., Tamura, M., Hirst, E., Harrison, D., 2009, MNRAS, 393, 229
\bibitem[\protect\citeauthoryear{Lyot}{1929}]{lyot29} Lyot, B., 1929, Ann. Observ, Paris (Meudon) 8, 1
\bibitem[\protect\citeauthoryear{Mancini et al.}{2016}]{mancini16} Mancini, L., Kemmer, J., Southworth, J., Bott,
K., Molliere, P., Ciceri, S., Chen, G., Henning, Th., 2016, MNRAS, 459, 1393
\bibitem[\protect\citeauthoryear{Madhusudham \& Burrows}{2012}]{madhusudham12} Madhusudham, N.
\& Burrows, A., ApJ, 747, 25
\bibitem[\protect\citeauthoryear{McCullough et al.}{2014}]{mccullough14} McCullough, P.R.,
Crouzet, N., Deming, D., Madhusudhan, N., 2014, ApJ, 791, 55
\bibitem[\protect\citeauthoryear{Mishchenko, Travis \& Lacis}{Mishchenko et al.}{2002}]{mishchenko02} Mishchenko, M.I., Travis,
L.D., Lacis, A.A., 2002, Scattering, Absorption and Emission of Light by Small Particles. Cambridge Univ.
Press, Cambridge 
\bibitem[\protect\citeauthoryear{Ordal et al.}{1988}]{ordal88} Ordal, M.A., Bell, R.J.,
Alexander, R.W., Newquist, L.A., Querry, M.R., 1988, Appl. Opt. 27, 1203
\bibitem[\protect\citeauthoryear{Pollack et al.}{1974}]{pollack74} Pollack, J.B. et al., 1974, Icarus,
23, 8
\bibitem[\protect\citeauthoryear{Pont et al.}{2013}]{pont13} Pont, F., Sing, D.K., Gibson, N.P.,
Aigrain, S., Henry, G., Husnoo, N., 2013, MNRAS, 432, 2917
\bibitem[\protect\citeauthoryear{Press et al.}{1992}]{press92} Press, W.H., Teukosky, S.A., Vetterling,
W.T., Flannery, B.P., 1992, Numerical Recipes, 2nd edition, Cambridge Univ. Press, Cambridge
\bibitem[\protect\citeauthoryear{Richard et al.}{2012}]{richard12} Rchard, C. et al., 2012, 
\jqsrt, 113, 1276
\bibitem[\protect\citeauthoryear{Russell}{1916}]{russell16} Russell, H.N., 1916, ApJ, 43, 173
\bibitem[\protect\citeauthoryear{Seager, Whitney \& Sasselov}{Seager et al.}{2000}]{seager00}
Seager, S., Whitney, B.A., Sasselov, D., 2000, ApJ, 540, 504
\bibitem[\protect\citeauthoryear{Schmid, Joos \& Tschan}{2006}]{schmid06} Schmid, H.M., Joos, F., Tschan, D., A\&A, 452,
657
\bibitem[\protect\citeauthoryear{Scott \& Duley}{1996}]{scott96} Scott, A., Duley, W.W., 1996,
ApJS, 105, 401
\bibitem[\protect\citeauthoryear{Sengupta}{2008}]{sengupta08} Sengupta, S., 2008, ApJ, 683, L195
\bibitem[\protect\citeauthoryear{Sill}{1972}]{sill72} Sill, G.T., 1972, Commun. Lunar Planet Lab., 9, 191
\bibitem[\protect\citeauthoryear{Sing et al.}{2016}]{sing16} Sing, D.K. et al., 2016, Nature,
529, 59 
\bibitem[\protect\citeauthoryear{Sonneborn}{1982}]{sonneborn82} Sonneborn, G., 1982, in Jaschek, M \& Groth,
H.-G. (eds), Be Stars, IAU Symp. 98, Reidel, Dordrecht, pp 493-495
\bibitem[\protect\citeauthoryear{Stam et al.}{2006}]{stam06} Stam, D.M., de Rooij, W.A., Cornet, G., Hovenier, J.W., 
2006, A\&A, 452, 669
\bibitem[\protect\citeauthoryear{Spurr}{2006}]{spurr06}
Spurr, R., 2006, \jqsrt, 102, 316.
\bibitem[\protect\citeauthoryear{Stamnes et al.}{1988}]{stamnes88} Stamnes, K., Tsay, S.C., Wiscombe,
W., Jayaweera, K., 1988, \ao, 27, 2502.
\bibitem[\protect\citeauthoryear{Tomasko \& Smith}{1982}]{tomasko82} Tomasko, M.G., Smith, P.H., 1982,
Icarus, 51, 65
\bibitem[\protect\citeauthoryear{Todorov et al.}{2014}]{todorov14} Todorov, K.O., Deming, D.,
Burrows, A., Grillmair, C.J., 2014, ApJ, 796, 100
\bibitem[\protect\citeauthoryear{van de Hulst}{1968}]{vandehulst68} van de Hulst, H.C., 1969, J. Comp. Phys., 3, 291
\bibitem[\protect\citeauthoryear{West \& Smith}{1991}]{west91} West, R.A., Smith, P.H., 1991, Icarus,
90, 330
\bibitem[\protect\citeauthoryear{Vestrucci \& Siewert}{1984}]{vestrucci84} Vestrucci, P., \& Siewert, C.E.,
1984, J. Quant. Spectrosc. Radiative Transfer, 31, 177
\bibitem[\protect\citeauthoryear{Wakeford \& Sing}{2015}]{wakeford15} Wakeford, H.R., Sing,
D.K., 2015, A\&A, 573, A122
\bibitem[\protect\citeauthoryear{Wiktorowicz}{2009}]{wiktorowicz09} Wiktorowicz, S.J., 2009, ApJ, 696,
1116
\bibitem[\protect\citeauthoryear{Wiktorowicz \& Matthews}{2008}]{wiktorowicz08} Wiktorowicz, S.J.
\& Matthews, K., 2008, PASP, 120, 1282
\bibitem[\protect\citeauthoryear{Wiktorowicz et al.}{2015}]{wiktorowicz15} Wiktorowicz, S.J. et al.,
2015, ApJ, 813, 48
\bibitem[\protect\citeauthoryear{Young}{1973}]{young73} Young, A.T., 1973, Icarus, 18, 564
\bibitem[\protect\citeauthoryear{Yurchenko et al.}{2014}]{yurchenko14} Yurchenko, S.N., Tennyson, J., Bailey, J.,
Hollis, M.D.J., Tinetti, G., 2014, PNAS, 111, 9379
\bibitem[\protect\citeauthoryear{Zhou et al.}{2013}]{zhou13} Zhou, G., Kedziora-Chudczer, L., Bayliss, D.D.R., Bailey, J., 2013,
ApJ, 774, 118
\bibitem[\protect\citeauthoryear{Zhou et al.}{2015}]{zhou15} Zhou, G., Bayliss, D.D.R., Kedziora-Chudczer, L., Salter, G., 
Tinney, C.G., Bailey, J., 2015, MNRAS, 445, 2746
\bibitem[\protect\citeauthoryear{Zugger et al.}{2010}]{zugger10} Zugger, M.E., Kasting, J.F.,
Williams, D.M., Kane, T.J., Philbrick, C.R., 2010, ApJ, 723, 1168
\end{thebibliography}







\bsp	
\label{lastpage}
\end{document}